\documentclass[twocolumn]{jpsj2} 
\usepackage{graphicx}

\newcommand{\lsco}{La$_{2-x}$Sr$_x$CuO$_4$}
\newcommand{\ybco}{YBa$_2$Cu$_3$O$_{7-\delta}$}
\newcommand{\ybcc}{YBa$_2$Cu$_4$O$_8$}
\newcommand{\bscco}{Bi$_2$Sr$_2$CaCu$_2$O$_{8+\delta}$}
\newcommand{\pcco}{Pr$_{2-x}$Ce$_x$CuO$_{4-\delta}$}
\newcommand{\plccox}{Pr$_{1-x}$LaCe$_x$CuO$_{4-\delta}$}

\newcommand{\msr}{$\mu$SR}

\title{Field-Induced Uniform Antiferromagnetic Order Associated with Superconductivity
in Pr$_{1-x}$LaCe$_{x}$CuO$_{4-\delta}$}

\author{
Ryosuke {\sc Kadono}\footnote{Also at The Graduate University for Advanced Studies (SOKENDAI)}, 
Kazuki {\sc Ohishi}\footnote{Present address: Advanced Science Research Center, Japan Atomic Energy Research Institute, Tokai, Naka, Ibaraki 319-1195}, 
Akihiro {\sc Koda}, 
Shanta R. {\sc Saha}\footnote{Present address: Department of Physics and Astronomy, 
McMaster University, Hamilton, Ont. Canada L8S 4M1},
Wataru {\sc Higemoto}$^1$,
Masaki {\sc Fujita}$^2$ and 
Kazuyoshi {\sc Yamada}$^2$}
\inst{Institute of Materials Structure Science, High Energy Accelerator Research Organization (KEK), Tsukuba, Ibaraki 305-0801\\
$^{1}$Advanced Science Research Center, Japan Atomic Energy Research Institute, Tokai, Naka, Ibaraki 319-1195\\
$^{2}$ Institute for Materials Research, Tohoku University, Sendai 980-8577}

\recdate{May 1, 2005}
\abst{Strong correlation between field-induced antiferromagnetic (AF) 
order and superconductivity is demonstrated for an electron-doped cuprate
superconductor, \plccox\ (PLCCO).  In addition to the specimen
with $x=0.11$ (which is close to the AF phase boundary, $x\simeq0.10$), we show that
the one with $x=0.15$ ($T_c\simeq16$ K at zero field) also exhibits the
field-induced AF order with a reduced magnitude of the induced moment.
The uniform muon Knight shift at a low magnetic field ($\sim10^2$ Oe) indicates that the 
AF order is not localized within the cores of flux lines, which is in a marked contrast
with theoretical prediction for hole-doped cuprates.  The presence of anomalous
non-diagonal hyperfine coupling between muons and Pr ions is also demonstrated
in detail.
}

\kword{electron-doped cuprates, superconductivity, magnetism, $\mu$SR\\}

\begin{document}
\sloppy
\maketitle


\section{Introduction}

The question whether or not the mechanism of
superconductivity in electron-doped ($n$-type) cuprates is common to
that in hole-doped ($p$-type) cuprates is one of the most interesting
issues in the field of cuprate superconductors,  which is yet to be answered. 
Theoretically, in the single band models such as the Hubbard model or
$t$-$J$ model\cite{Gross:87}, the electronic state of a hole on the CuO$_2$ plane
($3d^9\underline{L}$) is approximated by the Zhang-Rice
singlet between a local Cu$^{2+}$ spin and the hole on the oxygen atoms\cite{Zhang:88},
which is projected to the state of a spin-less hole ($3d^8$) on the copper ions.
This approximation makes the electronic state of holes and electrons virtually equivalent 
between $p$-type and $n$-type cuprates, leading to the prediction of
``electron-hole symmetry".   In reality, there is a limited number of 
compounds known to date as $n$-type, among which those belonging to the T' phase,
represented by Nd$_{2-x}$Ce$_x$CuO$_4$ (NCCO)\cite{Tokura:89,Takagi:89},
have been most actively studied. 
The T' structure is similar to that of the T phase represented
by \lsco\ (with reconfiguration of oxygen atoms) and thereby the latter is regarded as the
$p$-type partner.  Experimentally, it has been known for decades that 
the magnetic phase diagram as a function
of doping exhibits significant electron-hole asymmetry between those two
classes; while the antiferromagnetic (AF) order is strongly suppressed by small 
amount ($x\le 0.06$) of doping in $p$-type \lsco, the AF phase dominates over 
a wide region of doping (e.g., $0\le x\le 0.14$ in NCCO),  sharing a boundary 
with the superconducting phase\cite{Luke:90}.  In view of the $t$-$J$ model, 
it is argued that this asymmetry
can be understood by considering the difference in the actual electronic state
in such a way that the higher order hopping (transfer) terms are introduced to the
original models\cite{Tohyama:01}.

However, the recent observation of superconductivity in a new class of T' compounds,
La$_{2-x}RE_x$CuO$_4$ ($RE$ = Sm, Eu, etc., $T_c$ = 21--25 K)  suggests 
that there might be no such symmetry between $p$-type and $n$-type cuprates\cite{Tsukada:05}.
The new T' phase compound exhibits superconductivity without carrier doping 
(both rare earth 
and La ions are trivalent), which leads to a speculation that it might not be
crucial to have the AF insulator phase as a basis in modeling the $n$-type cuprates.  
Since the presence of the AF insulator (or, the Mott insulator) phase
is presumed to be an essential feature of $p$-type cuprates in a certain class of
theoretical models, the absence of the electron-hole 
symmetry would be a sign 
that the $n$-type cuprates may have their own mechanism for superconductivity
independent of the $p$-type partners.
As a matter of fact, there is increasing number of experimental evidence that
they are different in many respects. For example, the resistivity ($\rho$) in
the normal state of $n$-type cuprates follows a quadratic
temperature dependence ($\rho\propto T^2$) which is 
common to ordinary metals, whereas
a linear dependence is observed in optimally doped $p$-type cuprates.
This is further supported by the
recent NMR study of an $n$-type cuprate, \plccox\ (PLCCO, $x=0.09$), 
where the metallic Korringa law is observed upon the removal of 
superconductivity by applying
an external magnetic field above the upper critical field 
($H_{c2}\simeq50$ kOe)\cite{Zheng:03}.   Besides these, a
commensurate spin fluctuation is observed in the superconducting 
state of an $n$-type cuprate\cite{Yamada:03}, which is in marked contrast 
with the incommensurate spin fluctuation commonly found in $p$-type 
cuprates\cite{Cheong:91,Thurston:92,Dai:98,Arai:99,Hayden:04}.

The response to the external magnetic field is another potential clue to 
understand the ground state property of cuprates which are typical
type II superconductors.  They fall into a flux line lattice (FLL) state under a
magnetic field ($>H_{c1}$, the lower critical field),
where the superconducting order parameter is  locally suppressed in the 
center of flux (vortex cores).  In such a situation, the $t$-$J$ model predicts that the 
local carrier density would be partially proportional to
the order parameter so that the quasistatic AF correlation may develop in the
vortex cores\cite{Ogata:99}. A similar tendency is also predicted by the 
Hubbard model\cite{Takigawa:03} or that based on SU(5) symmetry\cite{Zhang:97}.
Interestingly, the recovery of quasistatic AF state under a moderate magnetic field
has been reported to occur in a whole variety of $p$-type cuprates including 
\lsco\ (LSCO)\cite{Katano:00,Lake:01,Lake:02,Kadono:04a,Savici:04}, 
\ybco\ (YBCO)\cite{Mitrovic:01,Miller:02}, \ybcc\ (Y$_{1248}$)\cite{Kakuyanagi:02}, 
and \bscco\ (BSCCO)\cite{Hoffman:02}.  
The primary issue in those results is whether or not the field-induced
AF state is localized in the vortex cores.  In this regard, while neutron diffraction is
not sensitive to the local structure over such a large length scale
(vortices with a core size of $\sim10^2$ \AA, separated
by $10^2$--$10^3$ \AA)\cite{Katano:00,Lake:01,Lake:02}, local spin probes like
nuclear magnetic resonance (NMR) and muon spin rotation (\msr) 
are sensitive to the local modulation of internal fields induced
by the AF cores. Indeed, although the case is not strong, there is certain evidence 
that such AF vortex cores may be 
realized\cite{Kadono:04a,Mitrovic:01,Miller:02,Kakuyanagi:02}.

In this paper, we present detailed report on the field-induced antiferromagnetism
in an $n$-type cuprate, \plccox, probed by \msr.  
A part of the work on a specimen with $x=0.11$ ($T_c\simeq26$ K) has been 
reported in the earlier paper\cite{Kadono:04b}, where we demonstrated 
that the response to external field 
is markedly different from that found in $p$-type cuprates; 
it is characterized by a uniform shift ($\Delta B\sim10^1$ Oe) 
of the internal field under an extremely low external magnetic field ($\sim10^2$ Oe).
More interestingly, $\Delta B$ exhibits unambiguous correlation with the
occurrence of superconductivity; it develops only below $T_c$ with an external field
lower than $\sim40$ kOe.
Here, we report new result on the specimen with $x=0.15$ ($T_c\simeq16$ K) 
which is situated deep in the superconducting phase from the phase
boundary to the AF phase ($x\simeq 0.10$).  While the observed response to
the external field is quite similar to that for $x=0.11$, the 
magnitude of the effect is considerably reduced.  These observations suggest
that the field-induced AF correlation coexists microscopically with superconductivity
in PLCCO. 

As demonstrated earlier in the specimen with $x=0.11$\cite{Kadono:04b}, 
the presence of non-diagonal hyperfine (HF) coupling parameter is confirmed in 
the new specimen by the muon Knight shift and susceptibility measurements 
with both field parallel and perpendicular to the $ab$-plane.
This establishes that muon probes the {\it in-plane} susceptibility 
of CuO$_2$ planes under a field {\it normal} to the planes, where
the Pr ions are coupled to the Cu ions by the superexchange interaction
and exert local field to muon throughout the non-diagonal HF coupling.
Thus, the occurrence of {\it in-plane} polarization associated with the 
field-induced AF phase leads to the shift of the internal field along $c$-axis,
which is mediated by the Pr ions. We show that this model provides
quantitatively consistent account of the field-induced antiferromagnetism
of CuO$_2$ planes observed by neutron diffraction in the 
specimen with $x=0.11$\cite{Fujita:04}.  
The success of the present model strongly suggests that the interpretation 
of data in the preceding work reporting a similar result in \pcco\cite{Sonier:03}
is inappropriate, leading to a highly overestimated size of Cu moments
in the field-induced AF state.

\section{Experimental}

\begin{figure}[hbt]
\begin{center}
\includegraphics[width=0.6\linewidth]{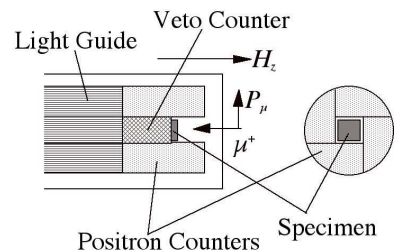}%
\caption{\label{fig1} A schematic view of experimental
set-up around the specimen for TF-\msr\ 
measurement, where $P_\mu$ is the initial muon polarization
and $H_z$ is the external magnetic field. The veto counter is used
to eliminate background signals from muons which missed the
specimen. Four positron counters are placed inside the cryostat
to minimize the positron path, which is crucial to attain a high
time resolution. }
\end{center}
\end{figure}

Single crystals of PLCCO with sizable dimensions were prepared by the 
traveling-solvent float zone method to obtain the specimen for 
$x=0.11$ ($\delta\simeq0.02$, $T_c\simeq26$ K) and $x=0.15$ 
($\delta\simeq0.05$, $T_c\simeq19$ K), the details of which have already been published 
elsewhere\cite{Fujita:03}. A large volume fraction 
and the sharp onset of  Meissner diamagnetism near $T_c$ 
(see below) demonstrated the high quality of these specimens.  
The \msr\ measurements were conducted on the M15 beamline at TRIUMF, Canada.
As illustrated in Fig.~\ref{fig1}, a slab of PLCCO crystal (measuring about 
5 mm$\times$8 mm$\times$0.5 mm) with the tetragonal $c$-axis perpendicular 
to the plane of the specimen was loaded onto a He gas-flow cryostat and a magnetic 
field (${\bf H}=(0,0,H_z)$) 
was applied  parallel to the $c$-axis (where $z\parallel c$).  Additional set of
measurements were made on the specimen with $x=0.15$
under a field parallel to the $ab$-plane.
In a transverse field (TF) geometry, the initial 
muon polarization was perpendicular to ${\bf H}$ so that the muon probed
the local field $B_z$ by spin precession at a frequency $\gamma_\mu B_z$
(with  $\gamma_\mu=13.553$ MHz/kOe being the muon gyromagnetic ratio).
Detailed zero-field (ZF) \msr\ measurements on the same specimen 
at various level of oxygen depletion indicated a weak random magnetism similar
to the case of \pcco, which is identified as being due to the small 
Pr moments\cite{Kadono:03}.

\section{Result and Discussion}

\subsection{Anomalous Muon-Pr Hyperfine Interaction in PLCCO}
\begin{figure}[hbt]
\includegraphics[width=0.9\linewidth]{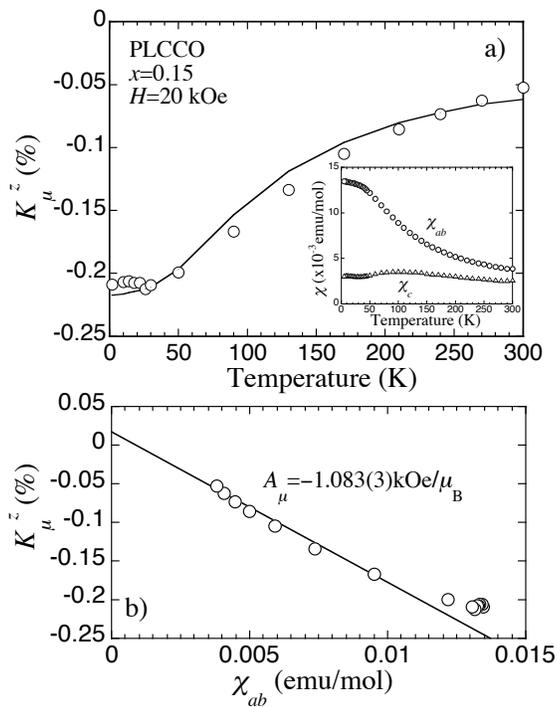}%
\caption{\label{fig2}   (a) The muon Knight shift with $H$ parallel to the $c$-axis, 
where the solid curve is proportional to $\chi_{ab}$.  (b) 
$K$-$\chi$ plot for the central frequency shown in (a). Inset: magnetic susceptibility 
vs temperature at $H$=20 kOe applied parallel  ($\chi_{c}$) or 
perpendicular ( $\chi_{ab}$) to the $c$-axis. }
\end{figure}

\begin{figure}[hbt]
\includegraphics[width=0.9\linewidth]{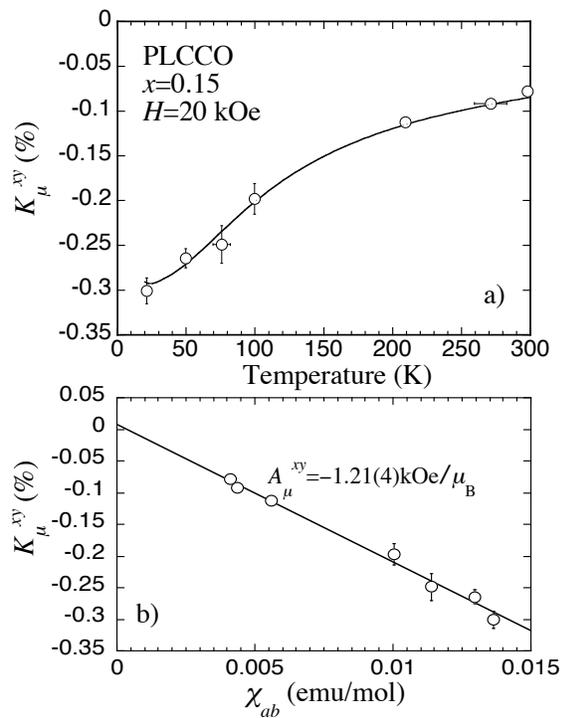}%
\caption{\label{fig3} (a) The muon Knight shift with $H$ parallel to the $ab$-plane, 
where the solid curve is proportional to $\chi_{ab}$.  (b) 
$K$-$\chi$ plot for the central frequency shown in (a).}
\end{figure}

The muon hyperfine parameter ($A_\mu$) is deduced from a 
comparison between the magnetic susceptibility ($\chi$) and 
the muon Knight shift ($K_\mu$) in the normal state. 
In the following, we assume that the crystal $c$-axis (= tetragonal axis) is always parallel
with the Cartesian $z$-axis, while the $ab$ direction is arbitrarily chosen in the $xy$ plane.  
Then, provided that the external field is parallel to the $c$-axis,
their relation in rare-earth metallic compounds is generally expressed as
\begin{equation}
K_\mu^z\simeq K_0+A_\mu\chi_{c}\frac{1}{N_A\mu_B}
=K_0+(A_{\rm c}+A^{zz}_{\rm dip})\chi_{c}\frac{1}{N_A\mu_B},\label{kzc}
\end{equation}
where $K_0$, $A_{\rm c}$ denote the respective 
contributions from the $T$-independent Pauli paramagnetism and from 
the polarization of conduction electrons by the Rudermann-Kittel-Kasuya-Yoshida 
(RKKY) interaction, $\chi_c$ denotes the susceptibility along the tetragonal axis, 
$N_A$ is the Avogadro number, $\mu_B$ is the Bohr magneton,
and $A^{zz}_{\rm dip}$ is the relevant component of the dipole tensor,
\begin{equation} 
A_{\rm dip}^{\alpha\beta}
=\sum_i \frac{1}{r^3_i}(\frac{3\alpha_i \beta_i}{r^2_i}-
\delta_{\alpha\beta}) \:\:(\alpha,\beta=x,y,z),
\end{equation}
which is predominantly determined by the nearest neighboring Pr ions
(with index $i$) at a distance ${\bf r}_i=(x_i,y_i,z_i)$ from the muon.  
Considering the earlier reports that the muon site is crystallographically unique 
and located near the oxygen atoms midway between the CuO$_2$ planes  
($z=0.25$ or 0.75 in the unit cell)\cite{Luke:90,Le:90}, we 
obtain a calculated value of $A_{\rm dip}^{\alpha\beta}$ as
\begin{equation}
A^{xx}_{\rm dip} = A^{yy}_{\rm dip}=-\frac{1}{2}A^{zz}_{\rm dip}=-482 \:{\rm Oe}/\mu_B,
\end{equation}
where the other non-diagonal terms are zero (i.e., $A^{\alpha\beta} = 0$ for $x\neq y\neq z$).
Assuming that the actual in-plane anisotropy is small (i.e., $A^{xx}_{\rm dip}\simeq
A^{yy}_{\rm dip}$), we have the Knight shift for the field parallel to the $ab$-plane, 
\begin{equation}
K_\mu^{xy}\simeq K_0+(A_{\rm c}+A^{xx}_{\rm dip})\chi_{ab}\frac{1}{N_A\mu_B},\label{kxyc}
\end{equation}
with $\chi_{ab}$ being the in-plane
susceptibility.
As shown in the inset of Fig.~\ref{fig2}(a), PLCCO exhibits a large anisotropy of $\chi$ 
between the in-plane ($\chi_{ab}$) and normal ($\chi_c$) directions, 
where $\chi_{ab}$ exhibits a significant increase with decreasing temperature 
while $\chi_c$ remains almost unchanged.  
This feature is least dependent on the Ce concentration
$x$, and well understood by the single-ion anisotropy of Pr ions 
(Pr$^{3+}$, $^3H_4$ multiplet),  considering
the tetragonal crystal electric-field and exchange interaction with 
Cu ions\cite{Sachidanandam:97}. 
The corresponding muon Knight shift versus temperature obtained for $x=0.15$
is shown in Figs.~\ref{fig2} for the case 
of ${\bf H}\parallel c$ (corresponding to $K^z_\mu$) and in Fig.~\ref{fig3} for
${\bf H}\parallel ab$ (corresponding to $K^{xy}_\mu$), where the external field is 20 kOe.
As evident in Figs.~\ref{fig2}(a) and \ref{fig2}(b), $K_\mu^z$ exhibits a
remarkable feature that it is mostly 
proportional to $\chi_{ab}$, which is in a stark contrast to  eq.~(\ref{kzc}) where $K_\mu^z$ is
expected to be linear in $\chi_c$.
This indicates that, while the muon Knight shift is primarily
determined by the hyperfine interaction with Pr ions, 
the effective field exerting on Pr ions is parallel to the
$ab$-plane under an external field applied parallel to the $c$-axis.
In contrast to the case of $K_\mu^z$, the in-plane Knight shift $K_\mu^{xy}$
exhibits a normal behavior; as shown in Fig.~\ref{fig3}(a) and  \ref{fig3}(b), 
it is proportional to $\chi_{ab}$ which is in line with eq.(\ref{kxyc}).

It is inferred from the Fourier analysis of the
TF-\msr\ spectra at higher external fields ($H\parallel c$)
that there are two symmetrical satellite peaks with respect to 
the central peak with about one half intensity
whose splitting becomes sufficiently large to be resolved 
above $\sim$25 kOe\cite{Kadono:03}.  
Provided that the rare-earth sites are randomly occupied by Pr and 
La ions, the observed spectral pattern suggests a binomial
distribution of hyperfine coupling constants.  Meanwhile,
our simulation indicates that such binomial distribution cannot be
reproduced by eq.~(\ref{kzc}) with the magnetic dipolar term as the 
predominant component.
Thus, we must introduce a Fermi contact-type non-diagonal 
hyperfine interaction, $A_f$, to account for the observed result, namely,
\begin{eqnarray}
K_\mu^z&\simeq & K_0+A_{f}\chi_{ab},\label{kzab}\\
K_\mu^{xy}&\simeq & K_0+(A_{f}+\overline{A}^{xy}_{\rm dip})\chi_{ab},\label{kxyab}
\end{eqnarray}
where we further assume that there is additional contribution of field-induced
(orbital) Pr moments along the $ab$-plane
to reproduce the difference between those two directions.
The $K$-$\chi$ plot for $\chi_{ab}$ in Figs.~\ref{fig2}(b) and \ref{fig3}(b) 
exhibits a linear relation with a small offset near the origin, from which we obtain
\begin{eqnarray}
A_\mu^z &=&\frac{dK_\mu^z}{d\chi_{ab}}=-1.083(3) \:{\rm kOe}/\mu_B = 
A_{f}\\
A_\mu^{xy} &=& \frac{dK_\mu^{xy}}{d\chi_{ab}}=-1.21(4) \:{\rm kOe}/\mu_B
= A_{f}+\overline{A}^{xy}_{\rm dip}.
\end{eqnarray}
 The value of $A_{f}$ is in reasonable agreement with that obtained 
previously, i.e., $ -969(2)$ Oe/$\mu_B$ in the specimen 
with $x=0.11$\cite{Kadono:04b}.  From the comparison with the calculated
value of the dipolar tensor, we have an estimated size of field-induced
Pr moment along the $ab$-axis at 20 kOe, 
$\mu_{\rm Pr}\simeq
(\overline{A}^{xy}_{\rm dip}/A^{xy}_{\rm dip})\mu_B=0.26(1)\mu_B$.
Here, we note that $\mu_{\rm Pr}$ is not a local spin but 
an effective dipole moment induced by the mixing of the $^3H_4$ multiplets under
an external field. Thus, it shrinks with decreasing field and thereby it is
consistent with the previous observation of fairly small Pr moments 
inferred from zero field-\msr\ study\cite{Kadono:03}.

It must be stressed that, while the origin of the above anomalous hyperfine
interaction is not clear at this stage, the non-diagonal term in eq.~(\ref{kzab}) 
plays a key role in probing the in-plane polarization of Pr ions which are directly
coupled to Cu spins in the CuO$_2$ planes.  In the following, we propose
a mechanism on how
the in-plane polarization in the CuO$_2$ planes exerts additional field
to muons sitting nearby the Pr ions.

\subsection{Field-Induced Antiferromagnetic Order  in the Superconducting State}

\begin{figure}[hbt]
\includegraphics[width=0.9\linewidth]{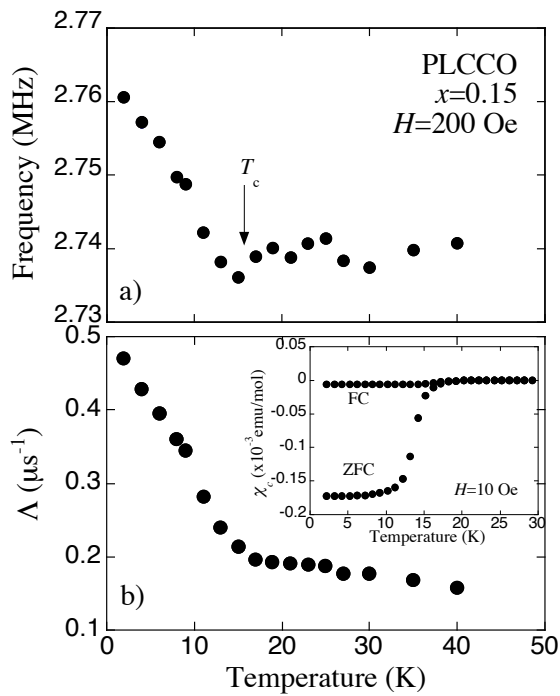}%
\caption{\label{fig4}
Temperature dependence of the muon precession frequency 
(a) and transverse spin relaxation rate (b) at $H_z=200$ Oe
(where $T_c\simeq 16$ K). }
\end{figure}

In the flux line lattice state of the type II superconductors, implanted muons 
provide a random sampling of the spatial field distribution $B({\bf r})$ so that the observed
\msr\ time spectrum is described by a complex polarization
\begin{equation}
P_x(t)+iP_y(t)=\int n(B)\exp(i\gamma_\mu Bt+\phi)dB,
\end{equation}
where $n(B)$ is the spectral density for $B({\bf r})$ characterized by a negatively shifted peak
corresponding to the van Hove singularity, and $\phi$ is the initial phase of muon spin 
rotation.\cite{Redfield:67}
Meanwhile, as shown in Fig.~\ref{fig4}, the \msr\ spectrum at $H_z=200$ Oe 
exhibits a large {\sl positive} shift of the peak frequency in the superconducting state of PLCCO,
which is apparently opposite to that associated with $n(B)$.  
Such a positive shift, or so-called ``paramagnetic 
Meissner effect", is occasionally found in the magnetization
of granular cuprate superconductors upon field-cooling (FC) with a small field ($\le1$ Oe), 
which has been interpreted as a manifestation of $d$-wave paring.\cite{Sigrist:95}
However, as shown in the inset of Fig.~\ref{fig4}, the possibility of attributing the
observed positive shift to the paramagnetic 
Meissner effect is ruled out by the magnetic response observed in both FC 
and ZFC (zero field-cooling) magnetization measurements; 
the bulk magnetization exhibits no anomaly to anticipate such a reversed magnetization.
The increase in the spin relaxation rate ($\Lambda$)
may be understood by considering the inhomogeneous magnetic field distribution due to
the FLL formation. 
These observations are common to the case with $x=0.11$, and 
similar to what has been observed in 
\pcco\ (PCCO)\cite{Sonier:03}.  

\begin{figure}[hbt]
\includegraphics[width=0.9\linewidth]{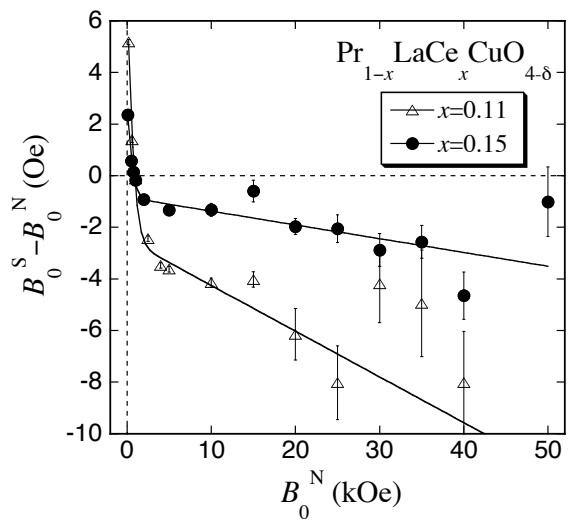}%
\caption{\label{fig5}Magnetic field dependence of the additional 
shift in the superconducting state,
where $B^S_0$ and $B^N_0$ correspond to the internal
field at 10 K and 35--40 K, respectively. Solid curve is a fitting
result by the model described in the text.}
\end{figure}

More surprisingly, this additional shift exhibits a strong and 
non-linear dependence on the external field.
The absolute magnitude of the shift, 
$\Delta B_z=B^S_0-B^N_0$ (frequencies divided by $\gamma_\mu$), is shown in 
Fig.~\ref{fig5}, where $B^S_0$ and $B^N_0$ denote the internal field felt by
muon at 10 K and at temperatures above $T_c$, respectively. It exhibits a steep decrease 
with increasing field, {\sl changing its sign} to negative above $\sim1$ kOe in both
cases of $x=0.11$ and 0.15.  
The field dependence of $\Delta B_z$ below $\sim1$ kOe indicates that
the change associated with the FLL state is completely
masked by the anomalous positive shift.
This is also consistent with a large magnetic penetration depth ($\ge3400$ \AA) 
reported for PCCO, which would lead to a shift with much smaller amplitude than 
the observed one (e.g., $\Delta B_z\sim-1$ Oe at $H=200$ Oe).
Above $\sim2$ kOe, $\Delta B_z$ is only weakly
dependent on the field up to $\sim40$ kOe, where the gradual increase
of the shift to the negative direction would be attributed to a small contribution
of bulk demagnetization effect.  

It might be argued that the observed behavior of $\Delta B_z$ below $\sim$1 kOe is 
qualitatively similar to that reported for PCCO ($H\le 2$ kOe)\cite{Sonier:03}.
However, while the latter suggests that $\Delta B_z$ approaches asymptotically 
to zero at higher fields, our data indicates that there is a change of sign in  
$\Delta B_z$; note that the lines in Fig.~\ref{fig5} representing the demagnetization effect
do not cross zero when they are extrapolated to lower fields.  
As discussed below, this is not understood by the simple model
proposed for the case of PCCO, but one has to consider a modulation (flip) of 
the hyperfine field $A^z_\mu$ induced by the external field.  
Here, it must be stressed that the hyperfine parameters reported in
the previous section is deduced from the data obtained at 20 kOe.
Unfortunately, $K^z_\mu$ in the normal state is too small 
to be measured at such low fields.
Provided that the change is described by 
the classical Langevin function, the entire field dependence including the flip is 
described in the following form,
\begin{equation}
\Delta B_z(H_z)=B^*\tanh(H_z/H_p)+c_dH_z+B_0,\label{bdif}
\end{equation}
where $B^*$ is the additional internal field induced by $H_z$, $H_p$ is the
characteristic field with which the sign of $A^z_\mu$ flips, $c_d$ is the
residual demagnetization term, and $B_0$ is the field offset.   The result
of fitting analysis using the above equation is shown by solid curves in 
Fig.~\ref{fig5}, where the obtained parameter values are listed in Table~\ref{tab1}.
The model reproduces data for both cases of $x=0.11$ and 0.15.

It is interesting to note that, apart from the steep change below $\sim2$ kOe
attributed to the flip of $A^z_\mu$,  the field dependence
of $\Delta B_z$ for $x=0.11$ is quite similar to that of 
field-induced Cu moments revealed by neutron diffraction
\cite{Fujita:04}; they exhibit least dependence on the field below 
$\sim$40 kOe.  This strongly suggests that the parameter $B^*$  in eq.~(\ref{bdif}), 
 which represents the amplitude of the internal field felt by muon, 
 is proportional to the Cu moments observed by neutron diffraction. 
Concerning the neutron result,  one may recall vigorous debate on the origin of
 field-induced effect observed in NCCO\cite{Kang:03}, whether 
 it is due to impurity phases \cite{Mang:03} or intrinsic to the CuO$_2$ plane.\cite{Matsuura:04}
 In PLCCO, the possible contribution of impurity phases
(Pr,Ce,La)$_2$O$_3$ has been investigated by the recent neutron diffraction 
experiment and they found no such effect below 70 kOe\cite{Kang:05}.

It is known in PLCCO that the Cu spins 
have a non-collinear AF structure, lying within the $ab$-plane\cite{Kang:05,Skanthakumar:93}.  
Provided that the field-induced AF phase has the same 
non-collinear structure, the $c$-axis component of the dipolar field from the
Cu moments would be zero.  Note, however, that this does not mean 
null shift, because the muon precession
frequency is determined by the vector-sum of dipolar fields, so that
the additional field along the $ab$-plane, $B_{xy}$, leads to a shift
\begin{equation}
\Delta B=\sqrt{H_z^2+B_{xy}^2} -H_z.
\end{equation}
Thus, one of the simplest models to explain
the positive shift at a low external field is to consider the effect of dipolar fields
along the $ab$ plane.\cite{Sonier:03}
In the case of PLCCO, 
the in-plane component of the dipolar field has two possible values,
$|B_{xy}|\simeq784$ Oe/$\mu_B$ or 365 Oe/$\mu_B$
(depending on the relative order of moments between two CuO$_2$ layers).
In order to explain the shift of $\sim$10 Oe at $H_z=200$ Oe
for $x=0.11$, we have to assume $|B_{xy}|\simeq 64$ Oe corresponding to
the Cu moment size of 0.1--0.2 $\mu_B$. Unfortunately, this is obviously too large to be
reconciled with the result of neutron diffraction, where the moment size of Cu ions 
induced by the external field is estimated as $\sim10^{-2}\mu_B$\cite{Fujita:04,Fujita:04-2}. 
Moreover, as pointed out
earlier, it does not explain the change in the sign of shift observed at higher fields.
Therefore, it is unlikely that the Cu moments directly 
contribute to $\Delta B_z$, in contrast to what has been suggested for the case of 
PCCO\cite{Sonier:03}. 

On the other hand, it is inferred from neutron diffraction studies that
there is a strong superexchange coupling between
Cu and Pr ions in $R_2$CuO$_4$ ($R$ = Nd, Pr)\cite{Matsuda:90}.  
More specifically,  about 0.08 $\mu_B$ of 
Pr moments is known to be induced by Cu moments with 0.4 $\mu_B$
in Pr$_2$CuO$_4$\cite{Matsuda:90}, having a 
non-collinear spin structure in both sublattices\cite{Skanthakumar:93}.
In a mean-field treatment, the Cu ions exert an effective magnetic field
on the Pr ions so that the moment size of the Pr ions is proportional
to that of Cu ions, namely,  
\begin{equation}
\langle M_{\rm Pr}\rangle\sim \chi_{ab}J\langle M_{\rm Cu}(H_z)\rangle,
\end{equation}
where $J$ is the Cu$^{2+}$-Pr interaction energy and 
$\langle M_{\rm Cu}(H_z)\rangle$ is the field-induced Cu moments\cite{Matsuda:90}.
This would lead to an additional field at the muon site, 
\begin{eqnarray}
\Delta B_z({\rm Pr}) & \simeq & A_{f}(H_z)\langle M_{\rm Pr}\rangle,\\
A_f(H_z) & \simeq & A_f^*\tanh(H_z/H_p).
\end{eqnarray}
Thus, $\Delta B_z({\rm Pr})$  is induced by Cu ions which are
polarized by an external field in the superconducting phase.
Here, it is obvious that the presence of non-diagonal hyperfine coupling $A_{f}$ 
between muons and Pr ions plays a key role. It serves as a mediator
between the {\sl in-plane} polarization of Pr ions and  the hyperfine field along the 
$c$-axis on the nearby muons. 
Provided that the magnitude of $A_{f}$  
at lower fields is close to that determined at 20 kOe ($\simeq1$ kOe/$\mu_B$), 
our estimation indicates that about $0.01\mu_B$ of Pr moments, 
which may be induced by $\sim0.05\mu_B$ of Cu moments, 
is sufficient to account for the amplitude of $\Delta B_z$ in the specimen with $x=0.11$.
A similar estimation yields $\sim0.02\mu_B$ of Cu moments for $x=0.15$.  It would be
due to this small moment size that the neutron diffraction 
were unsuccessful to detect the field-induced AF phase for the latter case\cite{Fujita:04}.

We would like to stress that the \msr\ results in the FLL state of $n$-type cuprates
(including those in PCCO) are qualitatively different from 
those obtained by similar techniques for the case of $p$-type cuprates in
two aspects. First of all, the effect of superconducting phase manifests itself in 
the well-defined frequency shift, whereas it is traceable 
only as an enhancement
of spin relaxation\cite{Savici:04,Mitrovic:01,Kakuyanagi:02} or
the change in the specific part of the field profile related to vortex cores\cite{Miller:02,Kadono:04a}
in $p$-type cuprates. 
This indicates that the induced 
polarization of Cu ions in $n$-type cuprates is quite uniform over the entire volume of 
the specimen, while it might be localized in the vortex core region in $p$-type
cuprates.  Secondly, the effect is highly nonlinear
to the external field, as demonstrated by the fact that
the strong influence of superconducting phase is observed
at an external field as low as $10^2$ Oe where the density of magnetic
vortices is very small (their distance being $\sim 4\times10^3$ \AA).  Considering
that most of the implanted muons are probing the region {\sl outside}
the vortex cores in the specimen at this low field range,  we conclude that the origin of
the enhanced frequency shift in the superconducting phase is not 
confined in the vortex cores. 
In addition to the case of $x=0.11$\cite{Kadono:04b}, 
the new result in the specimen with $x=0.15$ confirms that the 
small polarization of Pr ions (and of Cu ions) is present over the entire volume of 
the specimen under an external field, which on the other hand suggests that
the AF correlation becomes weak with increasing $x$.
We emphasize that this is the most important finding
of the present work, since it is only the local probes such as \msr\ that can 
investigate the homogeneity of the magnetic order in such a length scale.

As mentioned earlier, the theoretical models based on the strong 
electronic correlation (e.g., the $t$-$J$ model\cite{Ogata:99} or Hubbard
model plus superconducting correlation\cite{Takigawa:03}) 
predict that the AF correlation tends to precipitate in the vortex cores in the FLL state.  
The present result indicates that 
such situation is not realized in PLCCO, as demonstrated by the data 
at 200 Oe where the contribution of vortex cores is negligible.
On the other hand, it might be in favor of the 
quantum critical point scenario in understanding the competition between AF 
and superconducting phases\cite{Sachdev:03}.  
The observed effect in PLCCO might be understood by assuming that a parameter
controlling the quantum criticality between the AF and superconducting phase
is extremely sensitive to magnetic field.  Considering
the step-like response of CuO$_2$ plane to the external field, one may further speculate
that the parameter might be related with the time-reversal symmetry (TRS); 
needless to mention that TRS is broken by the application of an external field
(irrespective of its magnitude).

In summary, we have demonstrated in the single crystal specimens of PLCCO
with $x=0.11$ and 0.15 that, (i) the muon Knight shift along the tetragonal axis is 
proportional to the {\sl in-plane} susceptibility (perpendicular to the field
direction), and that (ii) it is strongly enhanced by the occurrence of superconductivity
with remarkable sensitivity to the low magnetic field.
The latter can be understood by considering the weak in-plane polarization of Pr ions
induced by field-induced Cu moments via superexchange interaction.
This also suggests strongly that there is a uniform weak polarization of 
Cu moments induced by an external field of 10$^2$ Oe, which occurs only in 
the superconducting phase of \plccox\ irrespective of FLL formation. 
This extraordinary sensitivity of CuO$_2$ planes 
to a magnetic field, developing exclusively in the superconducting state, will 
provide a strong criterion for identifying the true
electronic ground state of $n$-type cuprates.

\begin{acknowledgement}
We would like to thank the staff of TRIUMF for technical support
during the experiment. This work was partially supported by a Grant-in-Aid
for Scientific Research on Priority Areas and a Grant-in-Aid 
for Creative Scientific Research by the Ministry of Education, Culture,
Sports, Science and Technology, Japan. One of the authors (A.K.) was
supported by JSPS Fellowship.
\end{acknowledgement}

%




\newpage
\begin{table}
\begin{tabular}{c|cccc}
\hline
\hline
x  & $B^*$ (Oe) & $H_p$ (kOe) & $c_d$ & $B_0$ (Oe) \\
\hline
0.11 & $-9.6(2)$ & 1.04(3)  & $-0.13(2)$ & 6.9(2) \\ 
0.15 & $-3.6(1)$ &  0.80(4) & $-0.059(9)$ & 2.74(9) \\ 
\hline
\hline
\end{tabular}
\caption{\label{tab1} Fitting parameters obtained by analyzing data shown in
Fig.~\ref{fig5} using eq.~(\ref{bdif}).}
\end{table}

\newpage

\end{document}